\documentclass[12pt]{iopart}
\usepackage{graphicx}
\begin{document}

\title[Influence of tubular initial conditions on two-particle correlations]{
Influence of tubular initial conditions on two-particle correlations}

\author{R P G Andrade$^1$, F Gardim$^1$,
F Grassi$^1$,
Y Hama$^1$ and W L Qian$^2$}
\address{$^1$Instituto de F\'\i sica, Universidade de S\~ao Paulo, Brazil\\
$^2$Instituto de Ci\^encias Exatas, Universidade Federal de Ouro Preto, Brazil}

\ead{grassi@if.usp.br}
\begin{abstract}

A {\em unified}  
description of the near-side and away-side structures 
observed in two-particle correlations as function of $\Delta \eta-\Delta \phi$
is 
proposed for low to moderate transverse momentum. 
It is based on the combined effect of tubular initial conditions and 
hydrodynamical expansion. 
\end{abstract}


\section{Event-by-event 3+1 hydrodynamics}

\subsection{Overview of codes}

One of the most striking results in relativistic heavy ion collisions, at RHIC 
and the LHC,
 is the 
existence of  structures
  in the two-particle correlations 
plotted as function of the pseudorapidity difference $\Delta \eta$ and 
the angular spacing $\Delta \phi$.
These structures may have a common hydrodynamic origin:
the combined effect of
longitudinal high energy density tubes 
(leftover from initial particle collisions)   and 
transverse expansion.
In order to compare with the data mentioned above, event-by-event 3+1 hydrodynamics 
must be used. In this type of approach, 
some initial conditions are generated, an hydrodynamics code is 
run and results are stored; this is done many times thus 
mimicking experience.

 This method has been developed since 2001 by the brazilian collaboration
SPheRIO \cite{spherio,page} (some typical results
can be seen in \cite{sph1,sph2,sph3,jun}). 
It has been also studied since 2007 by 
of H.Petersen et al. \cite{urqmdhydro}. In 2010,
K.Werner et al. \cite{klaus1} and  B.Schenke et al. \cite{schenke}
have also started to use this method.
All groups except the last one assume an ideal fluid.
(Recently, various groups have also been  working on event-by-event 2+1 
hydrodynamics, see e.g. \cite{holopainen,qiu}.)

\subsection{Results on two-particle correlations}

\begin{center}
\begin{figure}[htbp]
\includegraphics[width=3.5cm]{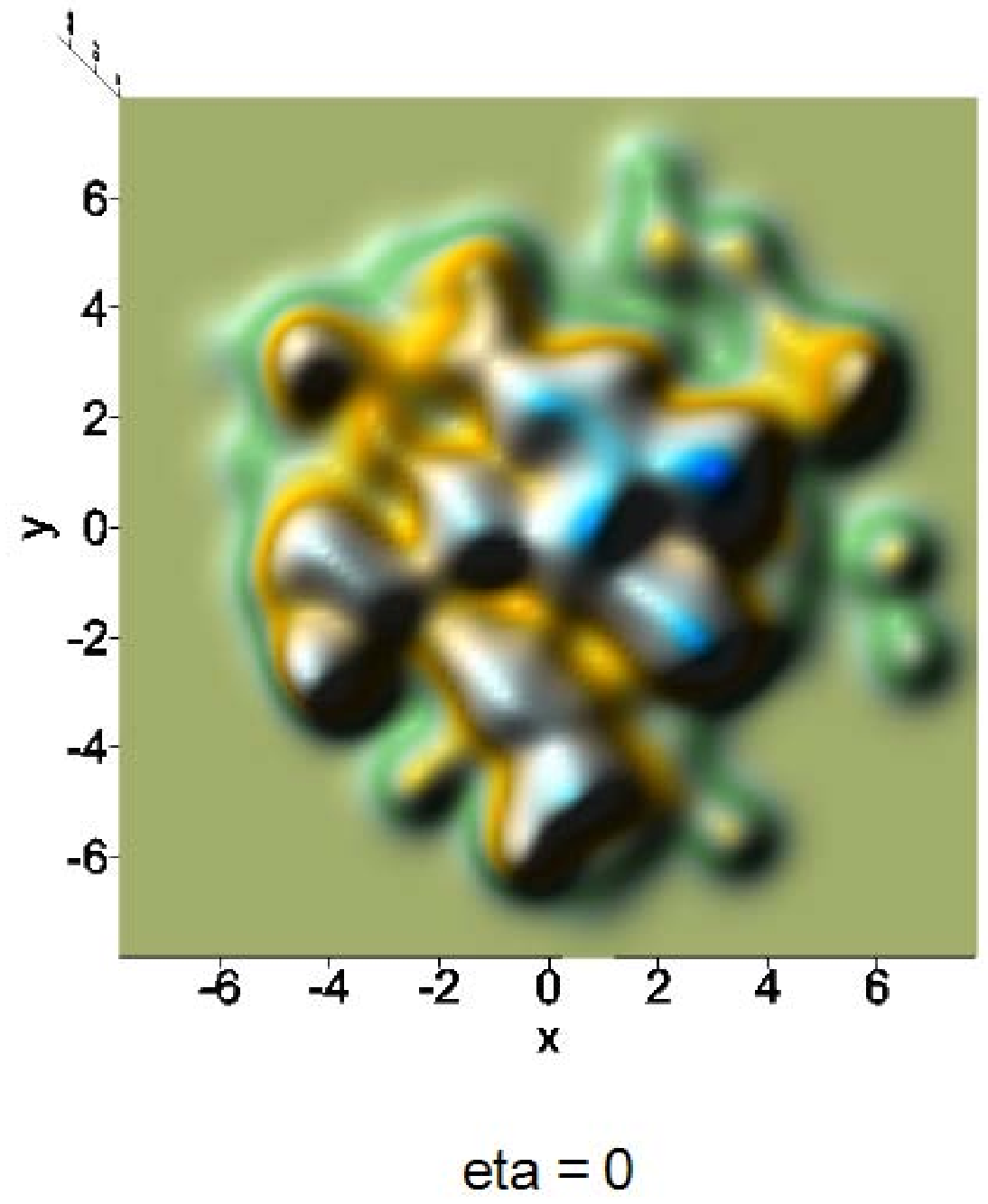}
\hspace*{2.cm}\includegraphics[width=4.cm]{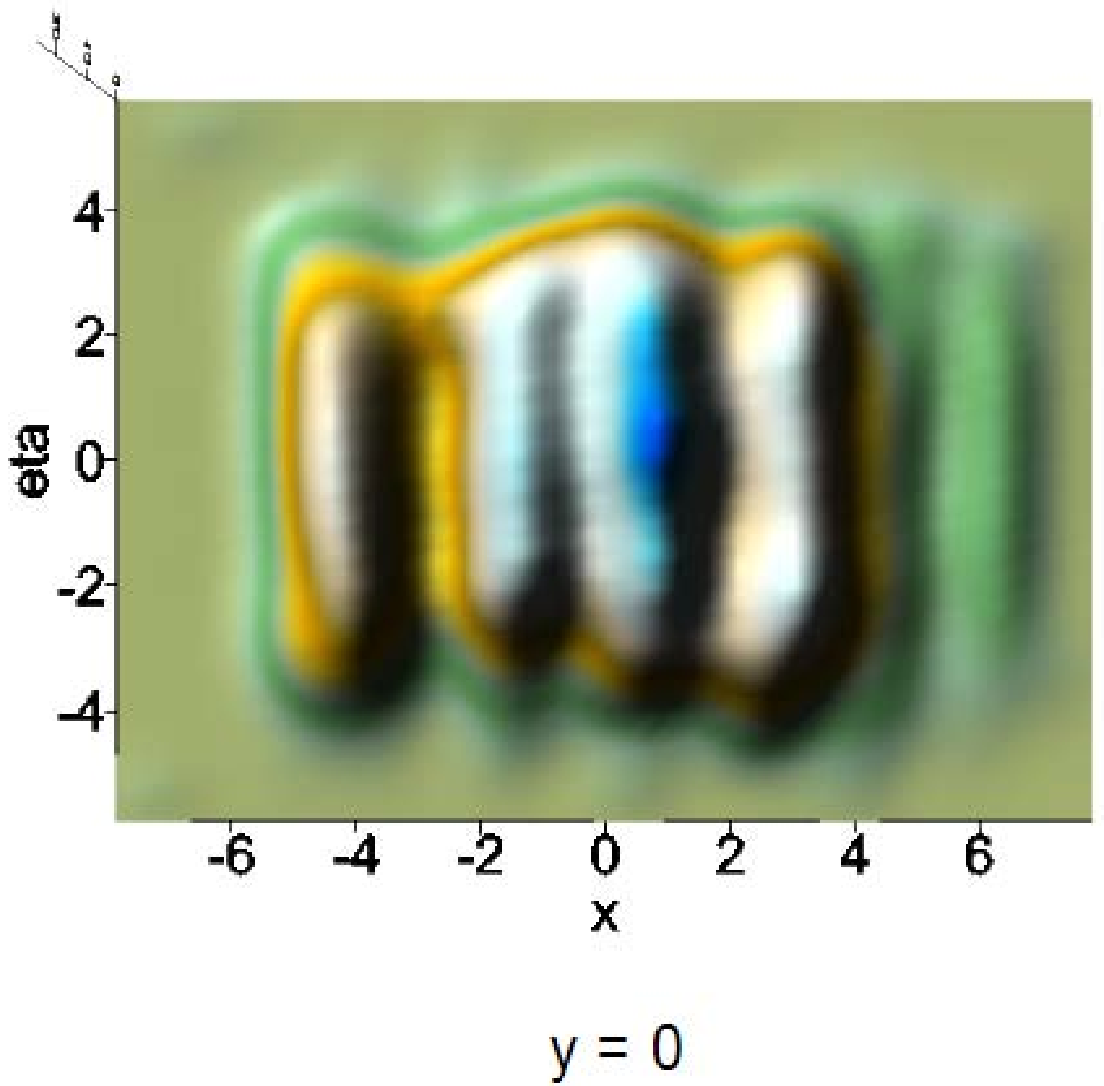}
\hspace*{2.cm}\includegraphics[width=4.5cm]{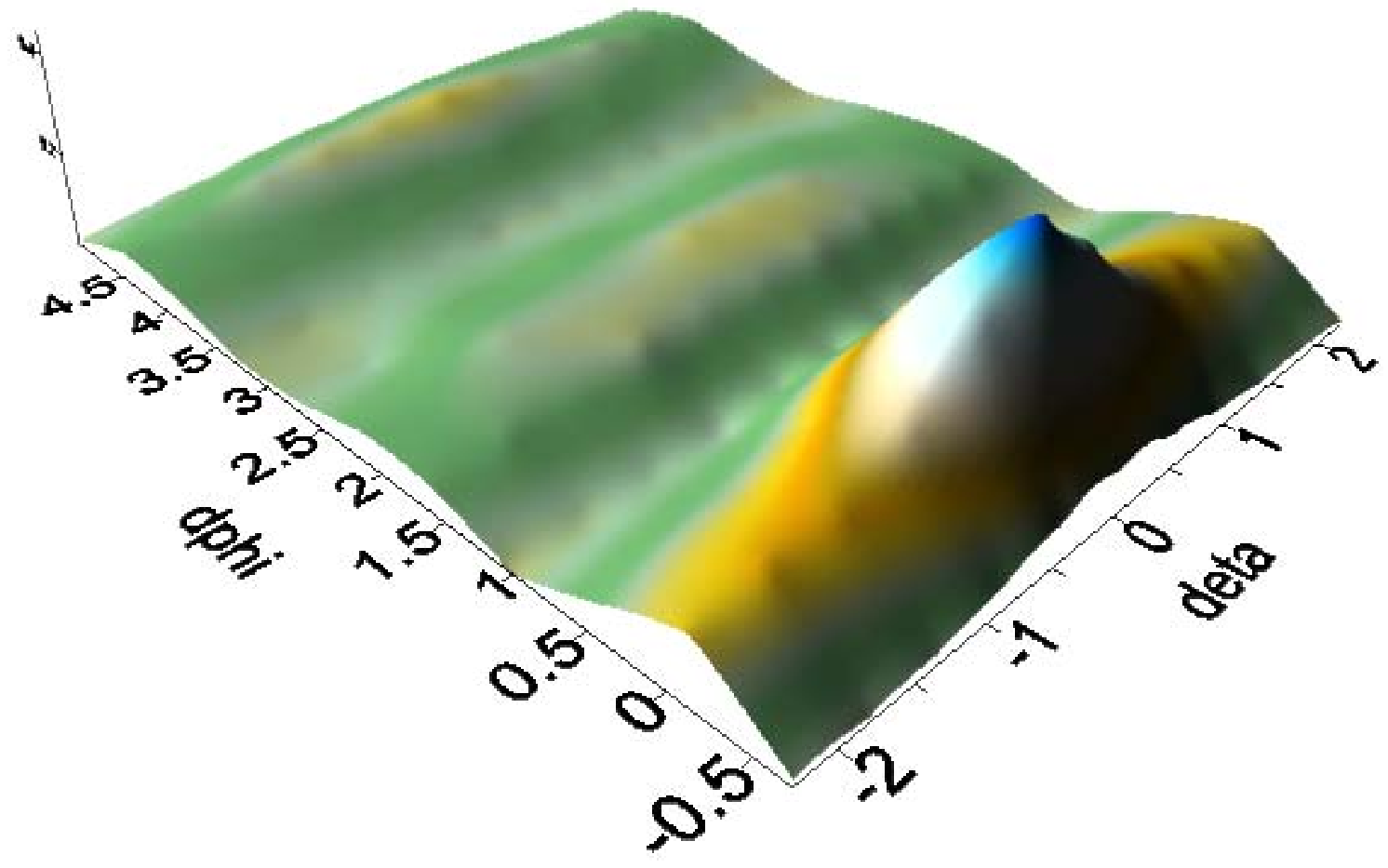}
\caption{\label{ic2d} NeXSPheRIO central Au+Au collisions  at 200 GeV A:
 initial energy density and example of resulting 
two-particle correlation.}
\end{figure}
\end{center}

In the 
NeXSPheRIO approach,
initial conditions, generated with the NeXus code,
have tubular structures and
 two particle correlations exhibit {\em near} and 
{\em away-side} ridges \cite{jun} as can be seen in figure \ref{ic2d}. In the calculation by
K.Werner et al.,
initial conditions, obtained with the EPOS code, 
also have tubular structures  and two particle correlations exhibit near-side \cite{klaus1} and 
{\em small} away-side ridges \cite{KWprivate}. 
(The other two groups mentioned above have no result on two-particle 
correlations.)

In addition to reproducing both the near and away-side structures,
NeXSPheRIO leads to good qualitative agreement with various data 
\cite{bnlart}:
1) for fixed $p_t^{trig}$ and increasing $p_t^{assoc}$, the near-side and away-side peaks decrease 
for central collisions
while for fixed $p_t^{assoc}$ and increasing $p_t^{trig}$, the peaks increase,
2) when going from central to peripherical collisions, the near-side ridge 
decreases and the away-side ridge changes from double to single hump,
3) 
for a mid-central window, the away-side 
ridge changes from single peak for in-plane trigger to double peak for out-of-plane trigger while for central collisions, it is always double-peaked.

\section{2+1 hydrodynamics: one tube model}

\subsection{Central collisions}

\begin{center}
\begin{figure}[htbp]
\includegraphics[width=4.cm]{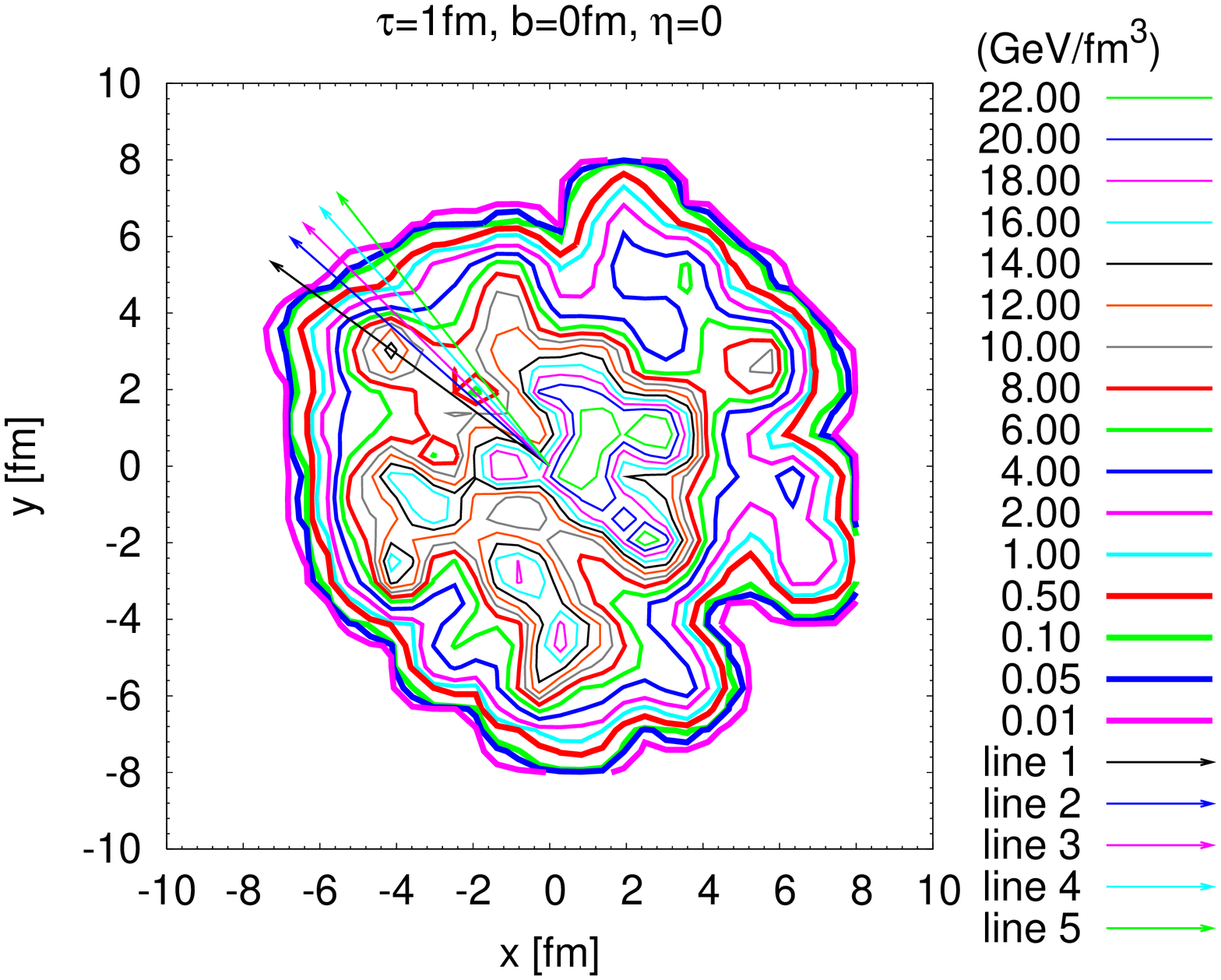}
\hspace*{1.5cm}\includegraphics[width=3.8cm]{fig2b.eps}
\hspace*{1.5cm}\includegraphics[width=4.cm]{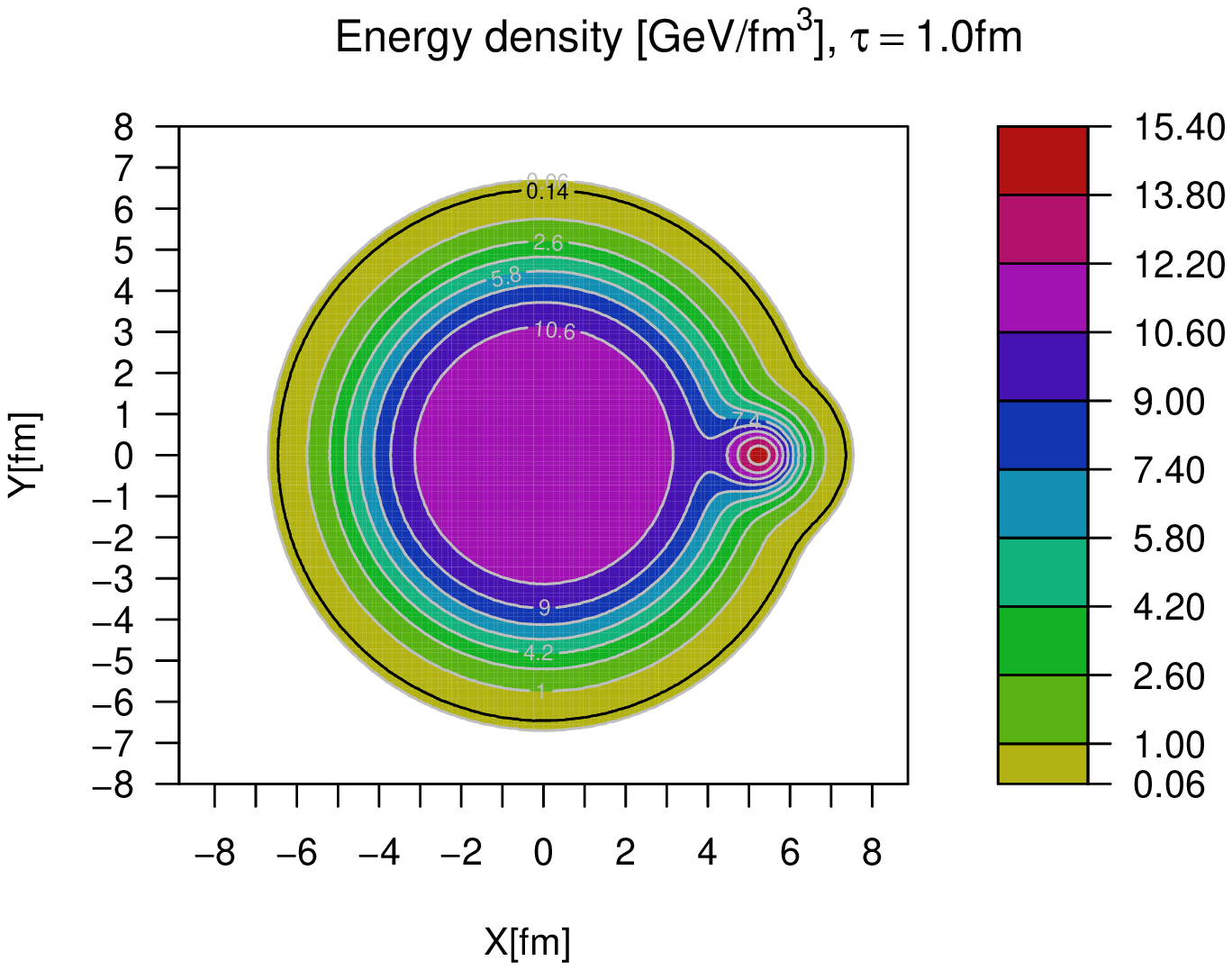}
\caption{\label{tube} Left: choice of a typical NeXus  tube
located near the border. Center: choice of the energy density profile for the tube and averaged background (solid lines). 
Right: resulting transverse slice of matter.}
 \end{figure}
\end{center}

When using NeXSPheRIO, it is not clear how the various structures
in the two-particle correlations are generated. To investigate this,
 we  study the transverse expansion of a {\em realistic} slice of matter with only
one tube (cf. figure \ref{tube}). Longitudinal expansion is assumed to be boost invariant and 
freeze out 
 to occur at some constant temperature.

\begin{center}
\begin{figure}[htbp]
\includegraphics[height=3.6cm]{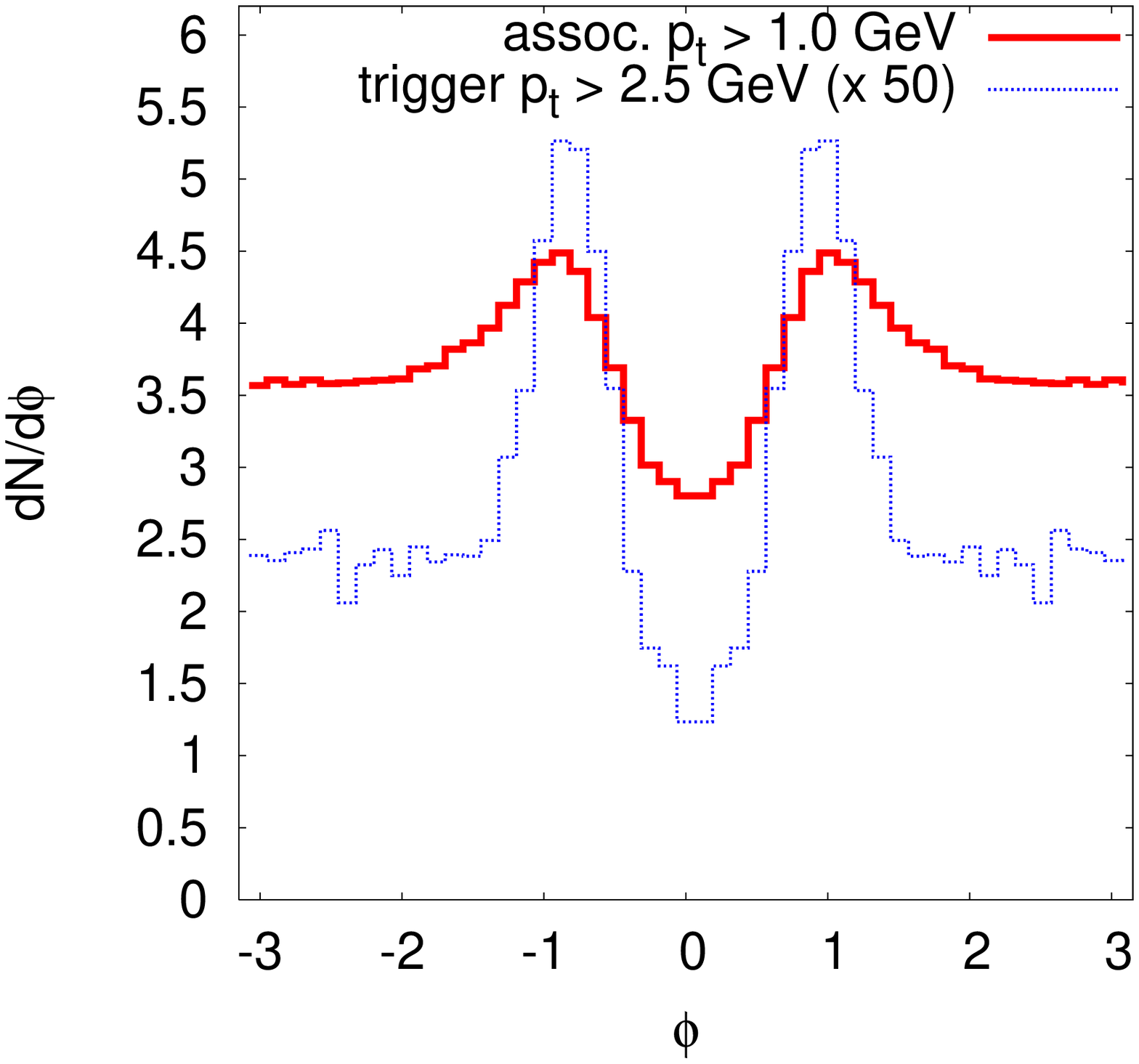}
\includegraphics[height=4.4cm]{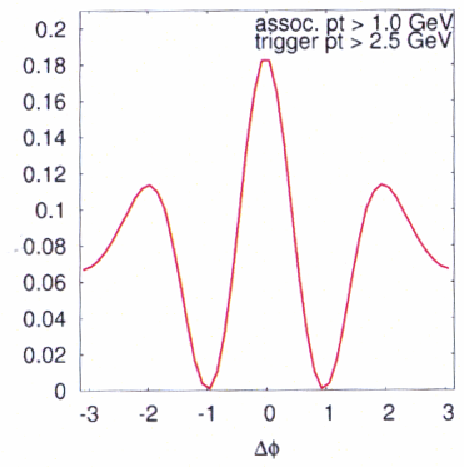}
\includegraphics[height=3.6cm]{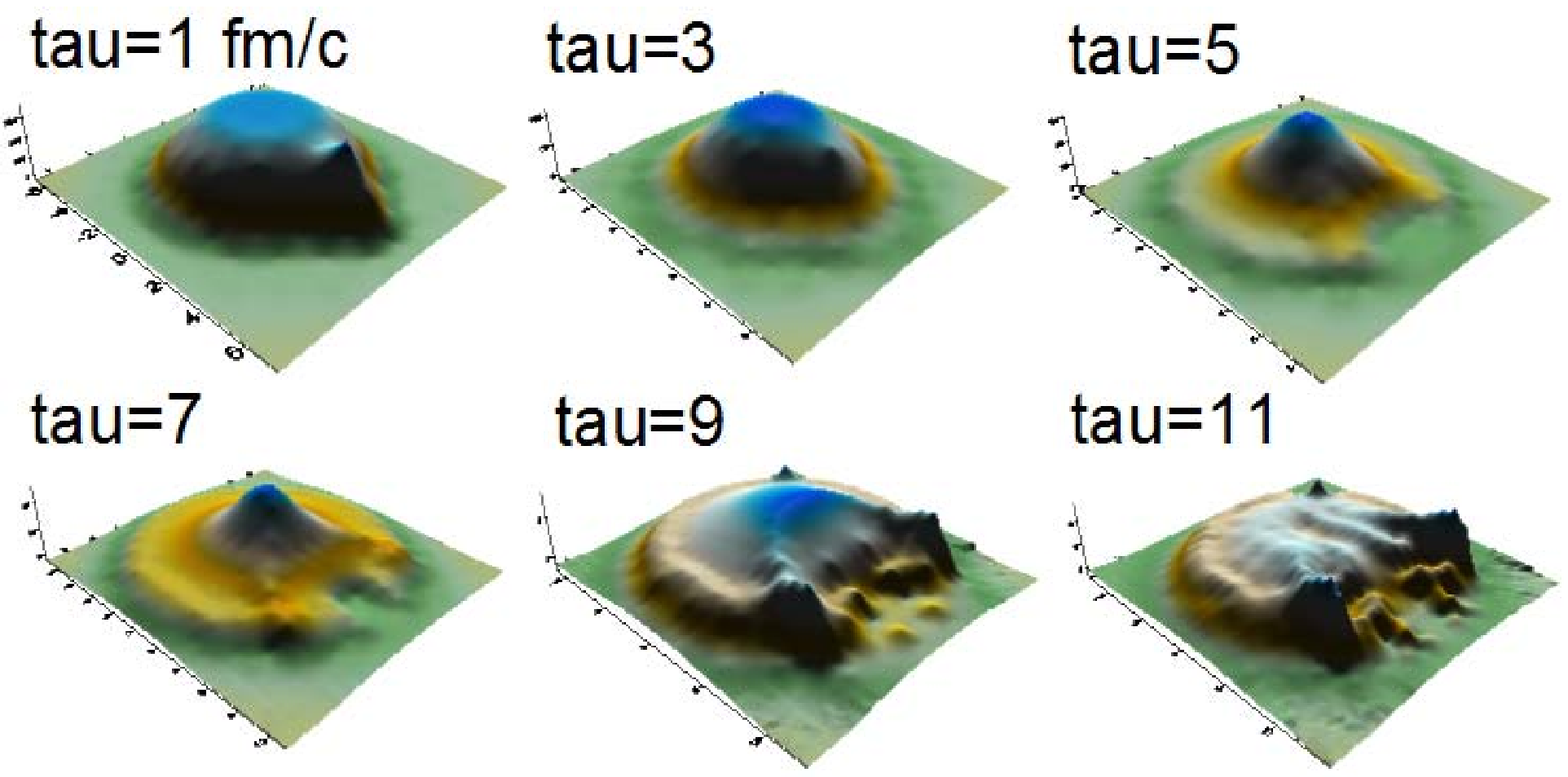}
\caption{\label{1dist} Left and center: single particle
angular distributions 
 and resulting two-particle correlation.
Right:  temporal evolution of the energy density in the slice. 
} 
\end{figure}
\end{center}

As seen in figure \ref{1dist},
the single particle angular distribution has 
two peaks located
on both sides of the position of the tube, more or less independently
of the transverse momentum value. 
 The peak spacing is $\Delta \phi \sim 2$ (this is not a parameter.)
The resulting two-particle 
angular 
correlation  has a large central peak at $\Delta \phi=0$ (corresponding to the
 near-side ridge)
and two smaller
peaks respectively at $\Delta \phi\sim \pm 2$ 
(associated to the double-hump ridge).
 We have checked  that
this 
 structure is robust by studying  the effect of the height and shape of the background, initial velocity, height, radius and location of the tube 
\cite{ismd09}.

  The occurrence of the two-peak emission  
can be understood from
 figure \ref{1dist}.
As time goes on, as a consequence of the tube expansion,
a hole appears at the location of the
high-energy tube (see also \cite{shuryak}).
This hole is surrounded by matter that piles up in a roughly semi-circular 
cliff of high energy density matter.
 The two extremities of the cliff emit
more fast particles than the background giving rise to the two-peaks in 
the 
 single-particle angular distribution.


\subsection{Non-central collisions}

We have also generalized this one tube model to non-central collisions \cite{rone}.
To get final results, one must sum on all tube positions.
For a given tube position, 
the 
single particle angular distribution is  more complicated (the double-peak
signature of the tube may 
be hidden by the elliptic flow) and
the shape of the two-particle angular distribution 
depends on the trigger angle with respect to the reaction plane.
After subtracting the
 elliptic flow  and averaging on the tube angular position,
 the two-particle correlation
is found to be
single peaked for in plane trigger and double peaked for out-of-plane trigger
in agreement with data.

\subsection{Effect of several tubes}

With these information, we can discuss what happens in a more complex event
 such as a NeXus event. 
Only the outer tubes need to be considered
and
will
 contribute with rather similar two-peak emission pattern at various angles 
in 
the single particle angular distribution. 
The two-particle correlation has a well-defined main
structure similar to that of a single tube surrounded by 
several other peaks and depressions due to trigger and associated particles coming from 
different tubes.  When averaging over many events they disappear and only the main
one-tube like structure is left.

\section{Summary}

Event-by-event 3+1
hydrodynamic  with
tubular initial conditions
 predicts two-particle correlations as function of
 $\Delta \eta-\Delta \phi$ in  
qualitative agreement with data.
Using a simpler model, we have seen that
near-side and away-side
structures
 are related to a
``two-horn'' emission from each tube.

It has been suggested \cite{triangular}
that both these near-side and away-side structures 
are due to the triangular flow ($v_3$)  produced by
a triangular shape (or triangularity $\epsilon_3$) present
in the initial conditions.
The one-tube model does have triangularity and triangular flow.
However, one can add three inner tubes to cancel $\epsilon_3$
and still have the same $v_3$ 
because only the outer tube contributes to this 
quantity. By removing this outer tube, leaving only the three inner tubes,
one can have 
$v_3\sim 0$ while  $\epsilon_3$ decreases a little but is non-zero.
Both $\epsilon_3$ (due to $r^3$) and $v_3$ are sensitive
to external tubes however $v_3$ more so. As a consequence, on an 
event-by-event basis
$v_3/\epsilon_3=cst$ will not be
 satisfied.
A more systematic study is in progress
(see also \cite{qin,qiu}).

We acknowledge funding from CNPq and FAPESP.


\section*{References}
\begin{thebibliography}{99}
\bibitem{spherio} Aguiar C E, Hama Y, Kodama T and Osada T 2001
{\em J.Phys. G} {\bf 27}  75; 2001 {\em J.Phys. G} {\bf 27} 551;
2002 {\em Nucl. Phys. A} {\bf 698} 639c 
\bibitem{page} $http://www.sprace.org.br/Twiki/bin/view/Main/SPheRIO$
\bibitem{sph1} Socolowski Jr O, Grassi F, Hama Y and Kodama T 2004 {\em 
Phys. Rev. Lett.} {\bf 93} 18230
\bibitem{sph2} Andrade R P G, Grassi F, Hama Y, Socolowski Jr O and
 Kodama T 2006 {\em Phys.  Rev.  Lett.} {\bf 97} 202302
\bibitem{sph3} Andrade R P G, Grassi F, Hama Y, Kodama T and Qian W L
2008 {\em Phys.  Rev.  Lett.} {\bf 101} 112301
\bibitem{jun} Takahashi J,  Tavares B M ,  Qian W L, 
Andrade R P G, Grassi F, Hama Y, Kodama T and Xu N
2009 {\em Phys. Rev. Lett.} {\bf 103} 242301
\bibitem{urqmdhydro} Petersen H, Steinheimer J, Burau G, Bleicher M,  
St\"ocker H 2008 {\em Phys. Rev. C} {\bf 78} 044901
\bibitem{klaus1} Werner K,  Karpenko Iu, Pierog T, Bleicher M and Mikhailov K
(2010)
 {\em Phys.Rev.C} {\bf 82}  044904
\bibitem{schenke} Schenke B, Jeon S and  Gale C 2011
{\em Phys. Rev. Lett.} {\bf 106} 042301 
\bibitem{holopainen} Holopainen H,  Niemi H and  Eskola K J 2011
{\em Phys. Rev. C} {\bf 83} 034901 
\bibitem{qiu} Qiu Z and Heinz Y arXiv:1104.0650.
\bibitem{KWprivate} Werner K {\em private communication}
\bibitem{bnlart}  Andrade R P G, Grassi F, Hama Y and Qian W L 2011
{\em Nucl Phys A} {\bf 854} 81
\bibitem{ismd09}  Hama Y, Andrade R P G, Grassi F and Qian W L 
2010 {\em Nonlin. Phenom. Complex Sys.} {\bf 12} 446
\bibitem{shuryak} Staig P and Shuryak E arXiv:1105.0676
\bibitem{rone} Andrade R P G, Grassi F, Hama Y and Qian W L
arXiv:1012.5275 
\bibitem{triangular} Alver B and Roland G 2010 {\em Phys. Rev. C} {\bf 81}
054905
\bibitem{qin} Qin G Y, Petersen H,  Bass S A and  M\"uller B
2010 {\em Phys. Rev. C} {\bf 82} 064903
\end{thebibliography}
\end{document}